# Elastic scattering of slow electrons by carbon nanotubes

M. Ya. Amusia[1, 2], A. S. Baltenkov[3]

[1] *Racah Institute of Physics, the Hebrew University, 91904, Jerusalem, Israel*
[2] *Ioffe Physical-Technical Institute, 194021, St. Petersburg, Russian Federation*
[3] *Arifov Institute of Ion-Plasma and Laser Technologies, 100125, Tashkent, Uzbekistan*

**Abstract**

In this paper we calculate the elastic scattering cross sections of slow electron by carbon nanotubes. The corresponding electron-nanotube interaction is substituted by a zero-thickness cylindrical potential that neglects the atomic structure of real nanotubes, thus limiting the range of applicability of our approach to sufficiently low incoming electron energies. The strength of the potential is chosen the same that was used in describing scattering of electrons by fullerene $C_{60}$. We present results for total and partial electron scattering cross sections as well as respective angular distributions, all with account of five lowest angular momenta contributions. In the calculations we assumed that the incoming electron moves perpendicular to the nanotube axis, since along the axis the incoming electron moves freely.

## 1. Introduction

The number of papers devoted to investigation of carbon nano-structures and their feasible applications rapidly increases (see, for example, [1-5] and references therein). The main elements of carbon nanostructures are graphene, fullerenes, endohedrals and carbon nanotubes. The first of these, graphene, is a crystalline structure formed by a single layer of carbon atoms assembled into a hexagonal lattice. Graphene can be considered as the building block of many carbon atoms nanostructures with tremendous potential for applications.

Fullerenes are graphene-type one-dimensional closed formations with an empty interior. Constructed of N carbon C atoms, they are denoted as $C_N$, of which best of all is studied $C_{60}$. The closeness of the shape of $C_{60}$ fullerene shell to a sphere results in its high stability and non-reactivity. that makes it possible to use their interiors as containers for atoms A. Compounds of this type were called endohedral atoms and labeled as A@$C_N$. The electrons of an atom A can be approximately considered as independent of the electron subsystem of $C_N$, and the real potential of the electron interaction with the fullerene shell can be replaced by a model potential.

Carbon nanotubes are tubes rolled up from graphene sheets. The length of such formations reaches tens of microns that are several orders of magnitude greater than their diameter, which is usually from one to several nanometers. It is natural to expect that the interaction of slow electrons with the nanotube wall is similar to the interaction of electrons with the same graphene sheet, that forms a fullerene $C_N$, which is very often described by a model potential.

Lohr and Blinder in [6] suggested the simplest of these model potentials. They assumed that the interaction of the atomic electron with the fullerene shell for electron wavelengths greater than the thickness of the shell can be described by a δ-type potential well which is different from zero within an infinitely thin spherical layer with radius *R* equal to the fullerene radius:

$$U(r) = -U_0\delta(r - R). \tag{1}$$

Bubble potential (1) is widely used to describe the processes of interaction of slow electrons with fullerenes and endohedral atoms [7]. Most attention was given to the potential created by $C_{60}$.

It is of interest to construct a model potential that describes the interaction between an electron and a carbon nanotube. In this article, based on the similarity in the nature of the interactions in fullerenes and nanotubes, we will suggest and investigate the model potential for the interaction of an electron with a nanotube, that is different from zero within an infinitely thin cylindrical layer and which by analogy with (1) can be called as bubble-tube potential.

The outline of our article is as follows. In the next Section 2, we will find regular and irregular solutions of the Schrödinger equation in circular cylindrical coordinates and adapt them to solving the problem of elastic scattering of a slow electron by a single cylindrical nanotube. By matching a regular solution of this equation inside a cylinder with a linear combination of regular and irregular solutions outside it, we derive a formula for phase shifts in the radial parts of continuum wave functions. These phases, as in the spherically symmetric case, determine the cross sections for elastic scattering of an electron by a cylindrical potential well. General formulas for the differential and total cross sections for elastic scattering of an electron in systems with cylindrical symmetry are derived in Section 3. Numerical calculations of the cross sections based on the derived formulas are carried out in Section 4. Final Section 5 contains our Summary and conclusions.

## 2. Wave equation

Consider the elastic scattering of an electron by a circular cylindrical potential well with radius $R$

$$U(\rho) = -U_0\delta(\rho - R). \tag{2}$$

In the Cartesian coordinate system, the axis of the potential cylinder will be assumed to coincide with the Z-axis. An electron with wave vector **k** falling on the cylinder will be assumed to be moving along the X-axis. In this case, the wave equation describing the motion of an electron in the XY plane in two-dimensional cylindrical coordinates $(\rho, \varphi)$ has the form

$$-\frac{\hbar^2}{2m}\left[\frac{\partial^2\psi}{\partial\rho^2} + \frac{1}{\rho}\frac{\partial\psi}{\partial\rho} + \frac{1}{\rho^2}\frac{\partial^2\psi}{\partial\phi^2}\right] + U(\rho)\psi(\rho,\phi) = E\psi(\rho,\phi) \tag{3}$$

Hereinafter, we use the same notations as in the article [8]. The variables $(\rho, \varphi)$ in Eq. (3) are separated when the wave function is substituted in (3) in the form

$$\psi(\rho,\varphi) = \psi(\rho)e^{im\varphi}. \tag{4}$$

We are interested in the continuous spectrum states, so we put $E = k^2/2$. Here and everywhere below, we use the atomic system of units. In them, Eq. (4) for $\psi(\rho)$ is written in the form

$$\psi'' + \frac{1}{\rho}\psi' - \left(2U(\rho) - k^2 + \frac{m^2}{\rho^2}\right)\psi(\rho) = 0. \tag{5}$$

Here, the primes denote differentiation with respect to the radius $\rho$. Going in (5) to the dimensionless variable $x = k\rho$, we rewrite the equation in the following form

$$\psi'' + \frac{1}{x}\psi' + \left(1 - \frac{2}{k^2}U(\rho) - \frac{m^2}{x^2}\right)\psi(x) = 0. \tag{6}$$

The primes here mean differentiation with respect to the variable $x$. The function $U(\rho)$ is equal to zero everywhere, except for the ε-neighborhood of the point $x_0 = kR$. Therefore, the equation (6) everywhere except for this point has the form of the Bessel equation [9]

$$\psi'' + \frac{1}{x}\psi' + \left(1 - \frac{m^2}{x^2}\right)\psi(x) = 0. \tag{7}$$

The general solution to this equation is a linear combination of the Bessel and Neumann functions [9]

$$\psi(x) = AJ_m(x) + BY_m(x). \tag{8}$$

The functions $J_m(x)$ and $Y_m(x)$ in (8) for large values of the argument behave like a sine and cosine

$$J_m(x) \approx \sqrt{\frac{2}{\pi x}}\sin\left(x - \frac{m\pi}{2} + \frac{\pi}{4}\right); \qquad Y_m(x) \approx \sqrt{\frac{2}{\pi x}}\cos\left(x - \frac{m\pi}{2} + \frac{\pi}{4}\right); \qquad x>>1. \tag{9}$$

Near zero ($x$ << 1), the Bessel function $J_m(x)$ is regular, and the Neumann function $Y_m(x)$ is singular.

Inside the cylinder, we represent the wave function of an electron in the form

$$\psi(x) = D_m J_m(x). \tag{10}$$

where $D_m$ is a $\rho$-independent coefficient to be determined. Outside the cylinder, this function, for the problem of electron scattering by a potential well (2), which we are solving, can be represented in the form of the following linear combination of the Bessel equation (7) solutions

$$\psi(x) = J_m(x)\cos\delta_m(k) + Y_m(x)\sin\delta_m(k). \tag{11}$$

Here, the functions $\delta_m(k)$ to be determined are phase shifts in the wave function of an electron that appear as a result of its passage across the potential well (2). They, as in the case of particle scattering in the spherically-symmetric potential, determine the cross section for elastic scattering of an electron by a target.

The wave function of the electron has to be continuous at the point $x_0 = kR$. Therefore, we have the following equality

$$D_m J_m(x_0) = J_m(x_0)\cos\delta_m(k) + Y_m(x_0)\sin\delta_m(k). \tag{12}$$

Hence, we obtain the expression for the coefficient $D_m$:

$$D_m = \cos\delta_m + \sin\delta_m \frac{Y_m(x_0)}{J_m(x_0)}. \tag{13}$$

The derivative of the wave function defined by equation (6) experiences a discontinuity at the point $x_0 = kR$. Indeed, integrating Eq. (6) in the $\varepsilon$-neighborhood of this point, we obtain

$$\int_{x_0-\varepsilon}^{x_0+\varepsilon} \psi'' dx = [\psi'(x_0+\varepsilon) - \psi'(x_0-\varepsilon)]_{\varepsilon\to 0} = -\frac{2}{k} U_0 \psi(x_0). \tag{14}$$

Thus, the jump in the logarithmic derivative of the wave function at the point $x_0 = kR$ is determined by the strength of the potential (2), i.e. by $U_0$ in equation (2)

$$\Delta L = \frac{1}{\psi(x_0)} [\psi'(x_0+\varepsilon) - \psi'(x_0-\varepsilon)]_{\varepsilon\to 0} = -\frac{2}{k} U_0. \tag{15}$$

Substituting into equation (14) the derivatives of the wave functions inside and outside the cylinder, we obtain the following equation

$$J'_m(x_0)\cos\delta_m + Y'_m(x_0)\sin\delta_m - D_m[J'_m(x_0) + J_m(x_0)\Delta L] = 0. \tag{16}$$

Inserting coefficient (13) into this equation and taking into account that the Wronskian of the Bessel equation (7) is $W(x) = Y'_m(x)J_m(x) - Y_m(x)J'_m(x) = -2/\pi x$ [9], we obtain, after elementary transformations of equation (16), the following expression for the phases of electron scattering on a cylindrical potential well (2)

$$\tan\delta_m = \frac{J_m^2(x_0)}{W(x_0)/\Delta L - J_m(x_0)Y_m(x_0)} \tag{17}$$

If the incoming electron moves not along the X axis orthogonal to the Z-axis, its wave function acquire a factor $\exp(ik_z z)$ that describes free motion along the Z axis, where $k_z$ is the electron linear momentum component along the nanotube axis. The expression (17) for $\delta_m$ remains the same, but the momentum $k$ is connected to the total incoming electron energy $E$ by the relation $E \equiv k^2/2 + k_z^2/2$, where $k^2/2$ is the energy of electron motion in the XY plane.

**3. Cross section of electron elastic scattering of a cylindrical potential (2)**

In the two-dimensional problem that we are considering, the wave functions of an electron with a certain projection of the angular momentum $m$ onto the X-axis have the form $Q_m(\rho)\exp(im\varphi)$, where the radial parts of these functions are

$$Q_m(\rho) = \begin{cases} D_m J_m(k\rho), & \rho \le R \\ J_m(k\rho)\cos\delta_m + Y_m(k\rho)\sin\delta_m, & \rho > R \end{cases}, \tag{18}$$

Radial functions (18) at large distances from the scatterer differ from the free motion functions only by the phase shift

$$Q_m(\rho) \approx i^m \sqrt{\frac{2}{\pi k\rho}} \sin\left(k\rho - \frac{m\pi}{2} + \frac{\pi}{4} + \delta_m\right), \tag{19}$$

where $\delta_m = \delta_{-m}$. Linear combination of these functions

$$\psi^+(\rho,\varphi) = \sum_{m=-\infty}^{\infty} e^{i\delta_m} Q_m(\rho) e^{im\varphi} \tag{20}$$

describes the process of electron scattering by a cylindrical potential well (2) and far from the scatterer, according to the book [10], is a superposition of plane and diverging cylindrical waves

$$\psi(x,\varphi) = e^{ikx} + f(\varphi)\frac{e^{ik\rho}}{\sqrt{-i\rho}}. \tag{21}$$

Here $\varphi$ is the angle between the X-axis and the direction of scattering in the XY plane; $f(\varphi)$ is the scattering amplitude, which in the two-dimensional case has the dimension of the square root of the length. The factor $-i = \exp(-i\pi/2)$ under the root is introduced to simplify the subsequent formulas. The scattering cross section per unit length of the cylinder $H$ along the Z-axis is

$$\sigma(k,\varphi) = |f(\varphi)|^2\, d\varphi. \tag{22}$$

It has the dimension of length. The scattering amplitude is

$$f(\varphi) = \frac{1}{i\sqrt{2\pi k}} \sum_{m=-\infty}^{m=\infty} (e^{2i\delta_m} - 1) e^{im\varphi}. \tag{23}$$

Integrating the square of the modulus of the amplitude over the scattering angle, we find the total cross section for elastic scattering of an electron by a potential well (2)

$$\sigma(k) = \int_0^{2\pi} |f(\varphi)|^2\, d\varphi = \sum_{m=-\infty}^{\infty} \sigma_m(k) \qquad \text{where } \sigma_m(k) = \sigma_{-m}(k) = \frac{4\pi}{k}\sin^2\delta_m. \tag{24}$$

The imaginary part of the scattering amplitude (23) at $\varphi = 0$ is related to the total scattering cross section (24) by the following relation

$$\operatorname{Im} f(0) = \sqrt{\frac{k}{8\pi^3}}\,\sigma(k), \tag{25}$$

That presents the optical theorem [10] for the two-dimensional case[1].

The results of numerical calculations of the cross sections for elastic scattering of electrons are presented in the next sections of this article.

**4. Numerical calculations of the scattering cross sections**

As a model problem illustrating the capabilities of the general formulas derived, we calculate the scattering cross section of a slow electron incident orthogonal to the axis of the cylindrical potential (2). Having in mind the similarity in the nature of electron interaction with graphene sheets forming both the $C_{60}$ fullerene and carbon nanotubes, we will suppose that the power of delta-potential $U_0$ in equations (1) and (2) is the same. Therefore, in the numerical calculations, we assume the value of $\Delta L$ in equation (15) to be equal to that calculated in [11, 12]

$$\Delta L = -2U_0 / k = -\beta(1+\coth \beta R)/k = -0.441/k\,. \tag{26}$$

Here $\beta = \sqrt{2I} \approx 0.441$ is the parameter in which $I$ is the electron affinity energy to the neutral $C_{60}$ fullerene: $I \approx 2.65$ eV. Radius $R \approx 6.64$ at. un. is the $C_{60}$ fullerene radius.

When calculating the phase shifts using formula (17), the regular Bessel function was defind by the formula

$$J_m(x) = \frac{1}{\pi}\int_0^{\pi} \cos(m\vartheta - x\sin\vartheta)d\vartheta\,. \tag{27}$$

Irregular at zero functions $Y_0(x)$ and $Y_1(x)$ were taken from tables contained in the book [9]. Functions with indices bigger than one were obtained using the recurrence relations

$$Y_{m-1}(x) + Y_{m+1}(x) = \frac{2m}{x}Y_m(x)\,, \tag{28}$$

applicable also to the regular Bessel functions.

The results of numerical calculations of the partial and total cross sections for elastic scattering of electrons by the bubble-tube potential well (2), carried out using formulas (24), are presented in Fig. 1. The results include the phases of the wave functions $\delta_m = \delta_{-m}$ with the projections of the angular moments $m$ = 0, 1, 2, 3 and 4. The scattering cross sections were calculated as a function of the electron linear momentum $k$. The radii of the potential cylinders (2) varied from $R$=1 to $R$=4 atomic units $a_0$ (Bohr-radius). The behavior of the curves in these figures is qualitatively similar. The partial cross section with $m$ = 0 dominates at small electron momenta. At low linear momenta, the curves with $m$ = 0 grow monotonically, almost linearly, and reach the maximal values, specified in the drawing panels. The upper frame of the figure cuts off the corresponding curves so that the behavior of curves with large quantum numbers $m$>0 can be seen in the figures. The total scattering cross sections in the panels of Fig. 1 represent curves oscillating around an imaginary monotonically decreasing curve.

The results of numerical calculations of differential scattering cross sections (22) are presented in Figure 2. The cylinder axis Z is perpendicular to the plane of the figure. The scattered beam of electrons with momentum $k$ moves along the X-axis. The curves shown in four panels of this figure represent the angular distributions of electrons after their elastic scattering by nanotubes with radii $R$ = 1, 2, 3 and 4 Bohr radius $a_0$. As in the calculation of the total scattering cross section in Fig. 1, in the calculations of the differential cross sections the contributions of nine scattering phases with $m$= 0, $m$= ± 1, $m$= ± 2, $m$= ± 3 and $m$= ± 4 are taken into account. Differential scattering cross section has the dimension of length/rad. Functions

$$B_m = \frac{1}{i\sqrt{2\pi k}}(e^{2i\delta_m} - 1) \tag{29}$$

in formula (23) are the coefficients of a linear combination of exponentials $\exp(im\varphi)$. These coefficients determine the shape of spectra $\sigma(k,\varphi)$ in Fig. 2.

[1] Note that in the book [10] there is a misprint in this formula, namely, it is printed $\pi$ instead of $\pi^3$.

## 5. Summary and conclusions

Here we propose a model for describing the process of interaction of electrons with cylinder-like carbon nanotubes. By analogy with the bubble potential for $C_{60}$ fullerene, we have proposed and discussed the Bubble-tube potential (2), which differs from zero only in an infinitely thin cylindrical layer in which carbon atoms placed. Obviously, such an idealization of the walls of carbon nanotubes is possible only for slow enough electrons, the wavelength of which significantly exceeds the real thickness of the walls, that is, the diameter of a carbon atom.

The idea of replacing the real potential with the model delta potential is not new. Apparently, it was first proposed in the classical article by Bethe and Peierls [13] in the study of interaction of nucleons in the deuteron. In this paper, it was shown that the behavior of nucleons outside the radius of their nuclear interaction can be described by imposing certain boundary conditions on the wave function of these particles, which is equivalent to introducing a delta-like pseudo-potential into the wave equation. The parameters of this zero-radius potential were determined in [13] from experimental data on the binding energy of nucleons in a deuteron.

When considering the interaction of slow electrons with the nanotube wall, we used an approach similar to [13]. We do not know the potential for the interaction of a carbon nanotube with an electron. However, based on the similaity of the nature of electron interaction with the surfaces of a nanotube and fullerene $C_{60}$, we can assume that the power $U_0$ of bubble-tube potential (2) and bubble potential (1) are equal. Note that in (1) the parameter $U_0$ is obtained from experimental data on the electron affinity for the $C_{60}$ molecule.

After the model potential of the nanotube was constructed, the Schrödinger equation was solved in cylindrical coordinates. It is shown that everywhere, except for points on the surface of the model cylinder, the solutions of the wave equation are the Bessel functions $J_m(kr)$ and $Y_m(kr)$. In the same way as in centrally symmetric problems with a deuteron [13] or fullerene $C_{60}$ [12], these were the spherical Bessel functions $j_l(kr)$ and $n_l(kr)$. Matching the wave functions on the cylinder surface leads to an equation for phase shifts in the radial parts of the wave functions of the continuum. These phases determine the cross sections for elastic scattering of electrons by nanotubes, which were calculated.

The results obtained here for incoming electrons that move orthogonal to nanotube's axis can be easily generalized to the case of arbitrary direction as it was mentioned at the end of Section 2.

It is obvious that the model developed here can be applied, for example, as well to the study of photoionization of atoms localized inside a cylindrical potential well [14], in order to detect the presence of confinement resonances in the cross sections of this process.

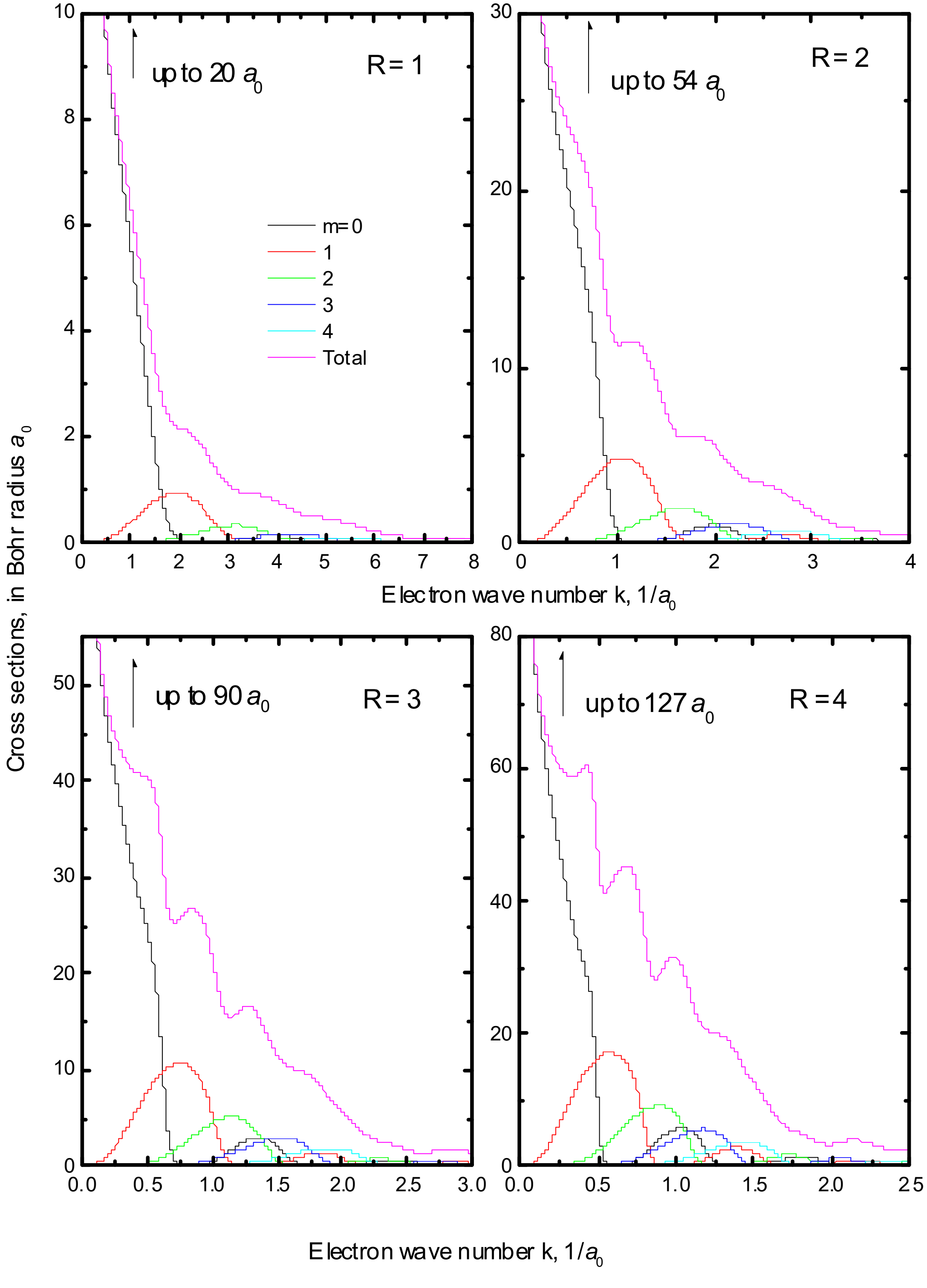

Fig. 1. Total cross sections of electron elastic scattering by the bubble-tube potential (2) as a function of electron wave number $k$. Curves with $m$ = 0 are the result of calculating the partial cross section with phase $\delta_0$; partial cross sections 1, 2, 3 and 4 are result of the calculation with the corresponding phase shifts; Total is the sum of all partial cross sections.

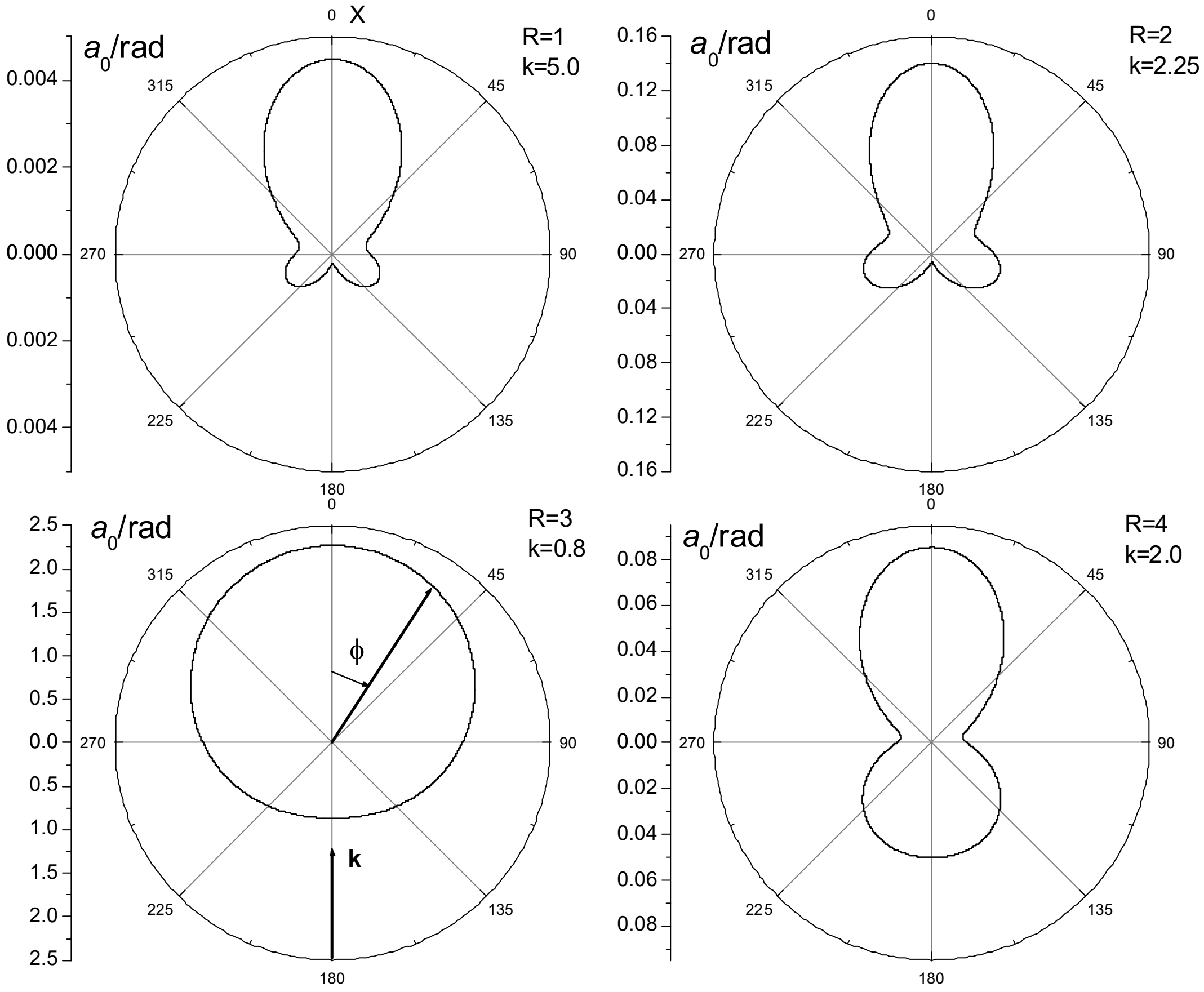

Fig. 2. Differential cross sections as function of polar angle $\varphi$ for cylinders radii $R$ = 1, 2, 3 and 4 $a_0$. Vector **k** is the direction of motion of the electron beam incident upon the cylinder, $\varphi$ is the angle of electron elastic scattering in the XY plane. The cross sections are calculated by formula (24) and are measured in units of $a_0$.